\date{}
\documentclass[reprint, amsmath,amssymb, aps]{revtex4-1}
\usepackage{amsmath}
\usepackage{amssymb}

\usepackage{comment}
\usepackage{booktabs}
\usepackage{graphics}
\usepackage{comment}
\usepackage{graphicx}
\usepackage{hyperref}
\usepackage{xcolor}

\def\equationautorefname~#1\null{Eq.\,(#1)\null}
\def\figureautorefname~#1\null{Fig.\,#1\null}
\def\sectionautorefname~#1\null{Sec.\,#1\null}

\definecolor{blue_cust}{HTML}{0066FF}
\definecolor{perfect_green}{HTML}{4FBF26}

\definecolor{crimson}{HTML}{DC143C}

\newcommand*\diff{\mathop{}\!\mathrm{d}} % ??
\begin{document}

%\bstctlcite{IEEEexample:BSTcontrol}

\title{Non-Analytic Behaviour in Large-deviations of the SIR model under the influence of Lockdowns}

\author{Leo Patrick Mulholland}
\email{leomulholland@hotmail.com}
\affiliation{School of Mathematics and Physics, Queen's University Belfast, Belfast BT71NN, United Kingdom}
\author{Yannick Feld}
\email{yannick.feld@uol.de}
\homepage{https://www.yfeld.de}
\author{Alexander K. Hartmann}
\email{a.hartmann@uol.de}
\affiliation{Institut f\"ur Physik, Carl von Ossietzky Universit\"at Oldenburg, 26111 Oldenburg, Germany}

\begin{abstract}
    We numerically investigate the dynamics of an SIR model with
    infection level-based lockdowns on Small-World networks.  Using a
    large-deviation approach, namely the Wang-Landau algorithm, we
    study the distribution of the cumulative fraction of infected individuals. We
    are able to resolve the density of states for values as low as
    $10^{-85}$. Hence, we measure the distribution on
    its full support giving a complete characterization of this
    quantity.  The lockdowns are implemented by severing a certain
    fraction of the edges in the Small-World network, and are
    initiated and released at different levels of infection, which are
    varied within this study.  We observe points of non-analytical
    behaviour for the pdf and discontinuous transitions for
    correlations  with other quantities such as the maximum fraction
    of infected and the duration of outbreaks.  Further, empirical rate
    functions were calculated for different system sizes, for which 
    a convergence is clearly visible indicating that the large-deviation principle is valid for
    the system with lockdowns. 
\end{abstract}

\maketitle

\date{\today}
%\author{Leo Patrick Mulholland$^{1*}$, Yannick Feld$^{2*}$ and Alexander K. Hartmann$^2$\\$^1$\emph{School of Mathematics and Physics, Queen's University Belfast, Belfast BT71NN, United Kingdom} \\$^2$\emph{Institut für Physik, Carl von Ossietzky Universität Oldenburg, 26111 Oldenburg, Germany}}

\section{Introduction}

The spread of infectious diseases is a phenomenon of great interest 
to many scientific fields \cite{Hethcote2000:MathematicsOfInfectiousDiseases,Pastor-Satorras2015:ZeroEpidemicThreshold:EpidemicsOnNetworks,Tang2020:reviewofinfectiousdisease}
and the recent outbreak of the SARS-CoV 2 pandemic has further increased this interest \cite{rahimi2021, hu2021, hu2022, cooper2020, dehning2020}.

The spreading of diseases can be modelled in a number of ways, such as deterministically with 
differential equations in the mean-field version of the \emph{susceptible-infected-recovered} (SIR) 
model \cite{Karmack1927:OriginalSIRModelPaper} or stochastically using methods such as agent models 
\cite{Bisin2022:agent,Bisin2022:agentlockdown,karaivanov2020social:agentmodel,Wu2022:AgentmodelWithLockdownAgentsCantMoveInLockdown, lorig2021}. 
A modelling more realistic than the mean-field is the study of the dynamics on 
networks \cite{Pastor-Satorras2015:ZeroEpidemicThreshold:EpidemicsOnNetworks,Du2021:NetworkmodelLockdownRemovingLongRange,plazas2021:ManySubNetworksforLockdownEdgeRemoval,Maheshwari2020:EdgeRemoval,Britton2016,ball2019stochastic}. 
Such methods can become arbitrarily complicated, such as layers of networks which 
represent different situations of contact \cite{Liu2018} or 
a time dependent network topology \cite{chen2014}.

In response to the SARS-CoV 2 pandemic, many governmental bodies 
imposed interventions to impede the spread of the disease \cite{brauner2021}. 
Thus, it has become rather fashionable to study the effect of 
\emph{disease prevention} methods \cite{Afshar2021:LockdownVaryingTransmission,Steffen2020:MaskAnalysisDifferentialEQ,Wu2022:AgentmodelWithLockdownAgentsCantMoveInLockdown,plazas2021:ManySubNetworksforLockdownEdgeRemoval,Priyanka2020,aguiar2021:multiscalenetwork,arazi2021,karaivanov2020social:agentmodel}.

Historically, great victories in the prevention of disease spread have been achieved through the distribution of \emph{vaccines} \cite{bonanni1999}.
Therefore it makes sense to include vaccines in the modelling \cite{Wang2016:Vaccines}.

However, the development and especially the approval of such pharmaceuticals can take a considerably 
long time \cite{buckland2005:vaccine_dev_process,Bok2021:CovidVaccine,Seunghoon2015:vaccinedev}. 
Consequently, the initial measures used to impede the spread of a previously unknown disease 
are so-called non-pharmaceutical methods \cite{mendez2021}. 
These include the wearing of face-masks in order to reduce the probability that a personal contact
 results in the spread of the disease, 
which may be modelled by reducing the transmission rate (or probability) 
dynamically \cite{Steffen2020:MaskAnalysisDifferentialEQ}.

Another important intervention is the imposition of \emph{lockdowns}. 
The idea is to greatly reduce the frequency of the contacts themselves. 
This can be modelled by reducing the transmission rate \cite{Afshar2021:LockdownVaryingTransmission} in mean-field and stochastic models, 
or by restricting the motion of walkers in agent based models \cite{Wu2022:AgentmodelWithLockdownAgentsCantMoveInLockdown,karaivanov2020social:agentmodel}. 
Network based models will typically model lockdowns through the removal or rewiring of 
edges \cite{Du2021:NetworkmodelLockdownRemovingLongRange,plazas2021:ManySubNetworksforLockdownEdgeRemoval,Maheshwari2020:EdgeRemoval,Britton2016,ball2019stochastic}.

In this work we are interested in the impact of lockdowns on the 
distribution of the cumulative fraction $C$ of infected individuals. In order to characterize $C$
comprehensively, i.e., obtain the 
probability distribution function (pdf) over its full support, we
extend the previous work on the large-deviation behaviour of SIR on networks \cite{feld2022:originalsirlargedevpaper, feld2023Vaccine, marks2023} 
by the inclusion of lockdowns. 

To our knowledge, no results are available in the literature in this regard. 
For this reason we keep the model relatively simple, i.e., 
we simulate a SIR model on networks from the Small-World 
ensemble \cite{Watts1998:OriginalSmallWorldPaper,Eubank2004:JustificationForUseOfSmallWorld}.

Technically, we  employ established
large-deviation
techniques \cite{bucklew2004:LargeDeviationBook,hartmann2002:largedeviationsalexprotiens,hartmann2014:largedeviationsalexising}. Note
that standard Monte Carlo sampling only allows access to the most
probable, i.e., typical regions.  Instead, by using a combination of
the Wang-Landau algorithm \cite{wang2001:originalwanglandaupaper} and
entropic sampling \cite{lee1993:originalentropicsamplingpaper}  we
are able to access the complete pdf  of $C$  which exhibits
probability densities as low as $10^{-85}$.  
Furthermore, we calculate the
empirical rate functions and verify whether the \emph{large-deviation
principle} holds 
\cite{den2000large,touchette2009,dembo2009largebook}.  
 Beyond the mere knowledge of
$P(C)$, having access also to the low-probability part allows us  to
comprehensively study correlations between different quantities  to
characterise the disease outbreaks.

This paper is ordered as follows: firstly, we recall the SIR model and introduce the
extension we use to model the lockdowns.  The parameters of
interest are discussed and the quantities we measure are defined.
Secondly, we discuss the ensemble of networks used.
In \autoref{sec:algorithms}, we present the implementation of the
simulation with the large-deviation methods.  Next the results of typical
Monte Carlo simulations that were used to find interesting points in the parameter
space are shown. For these points, we put forth the pdfs of the
fraction of cumulative fraction $C$ of infected and follow with characterizing
correlations pertaining to the disease-spread trajectories. We conclude with a
summary and discuss possible future directions.

\section{Model}

The basic dynamics of the disease spread are defined as follows: 
Each of the $N$ nodes in a connected network is assigned to one of the three states 
\emph{susceptible} (S), \emph{infected} (I) and \emph{recovered} (R). 
Here, our outbreak simulations begin at $\tau=0$ with five randomly chosen nodes 
assigned the infected state, while all other nodes are set to susceptible. 
Note that one could also start with one single initially infected, but
that would just increase the fraction of diseases which quickly die out,
which is not very interesting.

The states undergo a dynamical evolution at discrete times. For each
each time step $\tau$,  let $A_i$ be the number of infected neighbors
of node $i$.  If node $i$ is susceptible it will become infected at
time $\tau+1$ with the probability

\begin{equation}
    \lambda_i = 1 -
    (1-\lambda)^{A_i} \label{eq:infection_probability}\,,
\end{equation}
where $\lambda >0$ is the \emph{transmission probability} that a given
infected neighbour transmits the disease to $i$.  This is done for all
susceptible nodes.  Next, we iterate over all infected nodes that were not just infected in this 
time step and let each of them recover
at time $\tau+1$ with the \emph{recovery probability} $\mu>0$.  
These actions are repeated for time steps $\tau\to\tau+1$ until no infected nodes remain. 

Let $s(\tau)$, $i(\tau)$ and $r(\tau)$ be the fractions  of
susceptible, infected and recovered nodes, respectively,  at time
$\tau$.  Further, let the cumulative fraction of the infected be
$c(\tau) = i(\tau) + r(\tau)$.  The global properties of an outbreak
can be described by the final value of this quantity, i.e.,  
\begin{equation}
    C = \lim_{\tau \to \infty} \left(i(\tau) + r(\tau)\right)\equiv
    r(\tau=\infty),
\end{equation}
i.e., the total fraction of nodes that were infected at \emph{any} time
during the propagation.

The primary difference with the previous model is the incorporation of lockdowns: 
Once $i(\tau)$ reaches a threshold $\theta_l$, i.e., once $i(\tau) \geq \theta_l$, 
the disease does not continue to propagate on the original graph but on  the \emph{locked-down} graph instead. 
This is a subgraph of the original one and is obtained by randomly removing edges until a specified 
fraction $\eta$ (that we shall refer to as the \emph{severity} henceforth) 
of edges have been removed. 
Any lockdown can be lifted:
should the infection level $i(\tau)$  then fall below a second threshold $\theta_r < \theta_l$, 
the lockdown is released and the disease can propagate on the original network again. 
Note that the system, as in reality, may cycle in and out of lockdowns with multiple infection waves, 
until the propagation stops when the 
last node recovers. Note that the locked-down graph is calculated once per outbreak, i.e., at time $\tau=0$, so subsequent
lockdowns have the same underlying topology. One the other hand, when we average
over multiple runs a new locked-down graph is always created. 

%%%%%%%%%%%%%%%%%%%%%%%%%%%%%%%%%%%%%%%%%%%%%%%%%%%%%%%%%%%%%%%%%%%% continue below %%%%%%%%%%%%%%%%%%%%%%%%%%%%%%%%%%%%%%%%%%%%%%%%%%%%%%%%%5
\section{Ensemble}
We investigated Small-World ensembles \cite{Watts1998:OriginalSmallWorldPaper}, 
since real contact-networks between individuals have been observed to be modelled well by highly 
connected Small-World-esque networks \cite{Eubank2004:JustificationForUseOfSmallWorld}.

Technically, we initialise the graph with $N$ nodes in a ring structure, i.e., each node $i$ is first connected to its two 
subsequent neighbors $\{i,(i+1 \mod{N})\}$, $\{i,(i+2 \mod{N})\}$.
Each edge $\{i,j\}$ is then rewired with probability $p$ to a random node $j'$ resulting in the edge $\{i,j'\}$. 
These so-called \emph{long-range} edges introduce the Small-World characteristics of the network.  
We use $p = 0.1$ in this work. 
Should the resulting network be not connected, i.e., some nodes cannot be reached from others, 
we scrap the network and start the generation process afresh until a 
connected network is produced.

\section{Algorithms\label{sec:algorithms}}

\subsection{Outbreak Simulation}

To allow the sampling of very small probabilities using the large-deviation techniques, 
a method of manipulating the randomness of the disease outbreaks in a controlled manner is required. 
For a detailed explanation we refer to a previous publication \cite{feld2022:originalsirlargedevpaper} and 
just discuss the extension and main idea of the method here.

During a standard simulation the random numbers required to decide
whether nodes should become recovered or infected are  drawn on
demand, usually by calling a pseudo random number generator.  Nothing
prevents one, however, from drawing these random numbers beforehand
and storing them into, here, two random number vectors  $\xi_\lambda$ and
$\xi_\mu$, such that for each time step $\tau$ and node $i$ there is a
corresponding entry in the vectors.  Note that we require an estimate
of the maximum number of time steps $\tau_{\max}$  that the outbreak
is going to last for choosing an appropriate length of these vectors.
This will be discussed further in \autoref{sec:disease_duration}.

Additionally we store  
a vector $\xi_\eta$ containing all the edges of the investigated graph in randomized order.
Let $l$ be the total number of edges in the graph, i.e., the length of the vector $\xi_\eta$.
Then the pivot point $\theta_\eta = l (1-\eta)$, where $\eta$ is the fraction of edges affected by the lockdown,
 can be used to create the locked-down graph by using
the first $\theta_\eta$ edges of the vector. 
A list $\xi_P$ which contains the five initial infected nodes is also maintained.

Thus the outbreak and all measurable quantities are now deterministic functions 
of the randomness state $\Xi = \left(\xi_\lambda, \xi_\mu, \xi_\eta, \xi_P\right)$.

\subsection{Large-Deviation Sampling}\label{sec:large_Dev_sampling}

Our goal is calculating the complete probability distributions $P(E)$, 
where $E$ is some measurable quantity of our state $\Xi$, such as $C$ in our case, for a given 
network and specified parameters. For this purpose 
we have to employ methods more advanced than typical-event sampling to access states of particularly low probability 
in numerical simulations. 
Here, we use a large-deviation algorithm, employing a \emph{Markov chain  Monte Carlo} simulation 
\cite{newman1999barkema} of states given by the random numbers $\Xi$. The Markov chain 
$\Xi^{(0)}\to\Xi^{(1)} \to \ldots$ evolves by performing small changes to the given state $\Xi^{(t)}$.
The used Markov moves, which we have specifically tailored to 
our model, are now discussed.

With a probability of $1\%$, a \emph{lockdown move} is performed. 
First we decide how many edges we want to change by drawing a random integer between $1$ and $15$. 
For each edge we want to change we then randomly and uniformly choose an index $i\in [0,..,\theta_\eta)$ 
and $j\in [\theta_\eta,.., l-1]$ and swap the respective edges $\xi_\eta[i] \leftrightarrow \xi_\eta[j]$.

With probability $1\%$ a \emph{rotation} is performed. 
The elements of $\xi_\lambda$ and $\xi_\mu$ are shifted by $N$ elements to the right (50\%) or to the left (50\%) with periodic boundary conditions. 
This approximately reflects a shift of the trajectory by one time-step in either direction.
Note that this can be done very efficiently by not actually shifting the vectors in RAM but storing the 
offset instead.

With probability $7\%$ a \emph{patient move} is performed. 
There are two types of patient moves.
With a probability of $3/7$ a \emph{random patient move} is performed, where one of the entries of $\xi_P$ is 
redrawn by uniformly drawing a new node as initial patient. 
Note that duplicates are not allowed in $\xi_P$.
Otherwise, i.e., with probability $4/7$ a \emph{neighbour patient move} is performed.
One of the initial infected nodes is chosen randomly. Next one neighbour is chosen, each with probability $1/D$,
where $D$ is the maximum degree of the (not-locked-down) network. 
It is worth mentioning that for nodes that have less than $D$ neighbors
not choosing any neighbour is also possible. This ensure detailed balance, and thus,
in the long-time limit, all nodes are selected as being infected at the begin of the outbreak with the same probability.

With  the remaining probability of $92\%$, a \emph{standard move}
is performed, i.e.,
 changes are made to the values of the elements of $\xi_\lambda$ and $\xi_\mu$. 
Typically, with in 99\% of all standard-move cases, this is done as follows:
One of the vectors $\xi_\lambda$ and $\xi_\mu$ is chosen, and a random index $k$ and a random number $\chi \in [0,1]$ 
is drawn uniformly to set $\xi[k] = \chi$.
This choice of the vector and corresponding re-drawing of the random number is repeated $B$ times. 
The choice of $B$ does not determine the correctness of the algorithm, but rather the efficiency. The convention is to choose $B$ such that 
roughly $\sim50\%$ of trial configurations are accepted.

For the remaining 1\% cases of standard moves,
the random numbers corresponding to the infection and recovery of 
the initial infected as well as their immediate neighbours at $\tau = 0$ 
are all re-drawn uniformly from $[0,1]$. We observed that this strong special
 move improved the convergence of the Wang-Landau simulation.
 Since all of the random numbers are uniformly drawn, these moves do not skew any of the underlying statistics. 

The probability density functions (pdfs) are then calculated using the $1/t$ Wang-Landau algorithm \cite{Belardinelli-2007,Belardinelli-2008}, 
which is a variant of the original Wang-Landau algorithm \cite{wang2001:originalwanglandaupaper},
that does not suffer from the error-saturation problem
of the original algorithm \cite{Yan-2003, Belardinelli-2016}.

The basic idea is to initialise a non-normalised probability density estimate $P(C) = 1~\forall C$.
Then the a Markov-step is performed to generate a trial configuration $\Xi^\prime$ from the current configuration $\Xi^{(t)}$.
The configurations correspond to the cumulative number of infections 
$C^\prime=C(\Xi^\prime)$ and $C^{(t)}=C(\Xi^{(t)})$ respectively. They are 
used in the \emph{Metropolis-Hastings probability}
$\min\{1,P(C^{(t)})/P(C^\prime)\}$ to decide whether the 
trial configuration should be accepted, i.e., $\Xi^{(t+1)} = \Xi^\prime$ or rejected, i.e., $\Xi^{(t+1)} = \Xi^{(t)}$.
A multiplicative factor $f > 1$ is then used to update the distribution estimate, i.e., $P(C^{(t+1)}) = f P(C^{(t+1)})$,
all other entries, i.e., for $C\neq C^{(t+1)}$, are not changed.
The factor $f$ is iteratively reduced towards 1 
following some schedule and the pdf estimate converges
to the sought-after density function 
and just needs to be normalised in the end
by demanding $\int_0^1  P(C) \diff{}C= 1$.

One can split the support of $P(C)$, i.e., interval of allowed values of $C$,
 into multiple overlapping smaller
intervals and perform an independent simulation for each. This speeds up the simulation \cite{Schulz2003, landau2004}.
In the end one merges the obtained pdfs, using the fact that 
the pdfs have to match in the overlapping regions, at least within statistical fluctuations \cite{wang2001:originalwanglandaupaper, hartmann2002:largedeviationsalexprotiens}.

In this case, however, some sampling issues were encountered around a non-analytic point (``kink'') 
in the distribution when using 
multiple sampling intervals. 
This can be circumvented by making sure the kink is far from the interval boundaries,
but we mostly opted to just use one interval for the entire range and let the simulation run longer. 
Only for $N = 6400$, where a single interval required too much time for our taste, we used more intervals, six to be precise,
and applied sampling using \emph{Replica-Exchange}-Wang-Landau \cite{wang2001:originalwanglandaupaper,li2014new:REWL,vogel2014:rewl1,vogel2014:rewl2}, 
which is similar
to the described Wang-Landau algorithm but 
periodically tries to exchange configurations 
between overlapping intervals, hence the name \emph{Replica-Exchange}.

The pdfs were refined using entropic sampling \cite{lee1993:originalentropicsamplingpaper}, for details see 
Ref.~\cite{feld2022:originalsirlargedevpaper}.

\section{Simple Sampling}
In order to determine the points of interest in parameter space of the outbreak simulations, 
we perform some standard Monte Carlo simulations aiming at typical
outbreaks before running the large-deviation simulations.

\subsection{Transmission and Recovery Probabilities}

We want to study the behaviour of the model subject to lockdowns, so to obtain non-trivial
 results one should choose the parameters 
in such a way that the model would be 
in the pandemic phase if the lockdowns were absent. 
Working in discrete time, the parameters relating to disease spread are probabilities 
rather than rates as in the typical continuous time compartmental models. 
Following the previous work \cite{feld2022:originalsirlargedevpaper} of two of us, 
we set the recovery probability $\mu = 0.14$. The actual value is rather arbitrary 
and  basically sets the time scale.
What remains then is to choose the transmission probability. 
For the starting conditions with five initially infected individuals, 
we measured the epidemic threshold, without lockdowns, in the usual way by finding the value of
$\lambda$ which maximises the variance of $C$. We consider 
 increasing system sizes $N$ and perform a finite-size scaling analysis. 
This gives a critical transmission $\lambda_c(\infty) = 0.1186(5)$. 
Thus, we choose $\lambda = 0.2$, which comfortably places the system in the epidemic phase.

\subsection{Lockdown Parameters}
Having chosen our parameters pertaining to the spread of the disease itself, 
we need to choose the parameters governing the lockdown. 

When activating the lockdown, a fraction $\eta$ of edges,  is blocked, i.e., removed. The lockdown should
have a notable effect, so a natural choice is to thin out the edges to the
 \emph{percolation threshold} \cite{newman_book2010}  characteristic for the present Small-World ensemble. Clearly, one could also consider to remove
a smaller number of edges, which would change the behaviour
only slightly. Given the high numerical effort required, however, we concentrate on
the case where the lockdowns have the highest effect.
 Via finite-size scaling we measured the percolation threshold to be $\eta_c = 0.586(1)$,
which is consistent with previous results \cite{Newman1999:percolationThresh}. Thus, we used this value for the severity 
$\eta$.

As for the lockdown threshold $\theta_l$, which states the fraction of infected individuals above which the lockdown
is activated, and the corresponding release threshold $\theta_r$, we have chosen to typically use, 
unless stated otherwise, a  constant ratio $\theta_l = 8 \theta_r$.

\begin{figure}[htb]
    \centering
    \includegraphics[width = \linewidth]{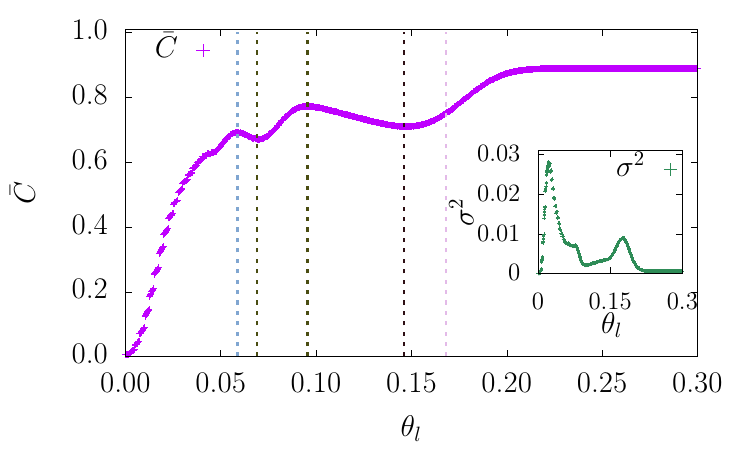}
    \caption{The average cumulative fraction of infected $\bar{C}$ as a function of the lockdown threshold 
    $\theta_l = 8 \theta_r$ for $N = 3200$, $\eta = 0.586$, $\mu = 0.14$, $\lambda = 0.2$. 
    The inset shows the variance. Error bars are smaller than symbol sizes and therefore omitted. 
    The dashed vertical lines indicate the points of interest, namely (in order) $\theta_l \in \left\{0.0588,0.0692,0.0955,0.1460,0.1683 \right\}$.}
    \label{fig:ScanThresh}
\end{figure}

Using these choices, we measured the average cumulative fraction $\bar{C}$ of infected 
as a function of $\theta_l$ for increasing system sizes. 
Each data point is averaged over 100,000 networks. Errors are calculated using bootstrap resampling 
\cite{Efron1979:bootstrap}. 
An example of such a behaviour for $N = 3200$ is shown on \autoref{fig:ScanThresh}. 
Results for other system sizes look similar.

%/Project/data/SimpleSamples/SmallWorld/NicePlots/ScanThresh

Intuitively, increasing the lockdown threshold will increase the cumulative fraction $C$ of infected. 
In particular, locking down too late can be seen to have no effect in containing the disease,
which also makes sense as the lockdown threshold needs to be reached for the lockdown to have any effect. 
Still, the behaviour is more complex, as we discuss next.

First, note that the presented result is obtained at the maximum resolution with respect to the possible
values of $\theta_l$, i.e., with increment $1/N$. 
Since the release-threshold is at one-eighth the lockdown-threshold, the release threshold 
only changes in every eighth data point, due to the integer nature of the measured quantities, 
giving rise to the apparent discontinuities in the early part of the curve.

Clearly, for small lockdown thresholds the lockdown is able to greatly contain and slow down the disease. 
Here, increasing the threshold leads to an increasing number of infections,
giving rise to a peak in the variance, see inset of \autoref{fig:ScanThresh},  around $\theta_l = 0.02$ indicating the transition to the 
epidemic phase. 

\begin{figure}[htb]
    \centering
    \includegraphics[width = \linewidth]{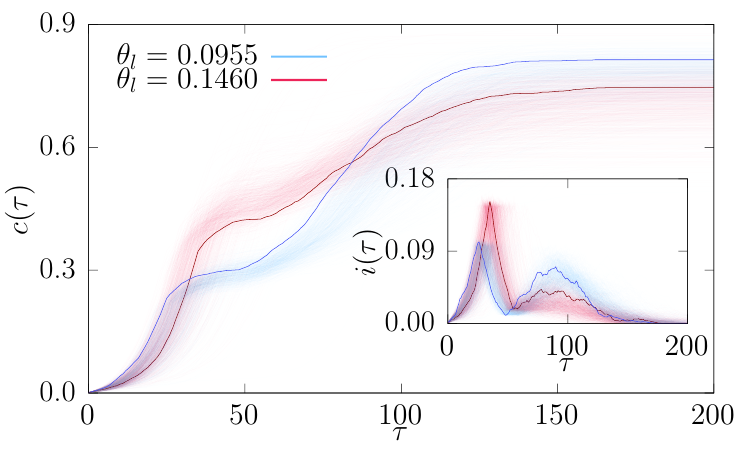}

    %\includegraphics[width = \linewidth]{Figures/plotcurves_latex_e_i_inset.jpg}
    %path in directory: SimpleSamples/SmallWorld/NicePlots/ScanThresh/NiceCurves/plotcurves_latex_e_i_inset.gp
    \caption{Fraction of cumulative fraction $c$ of infected  as a function of time $\tau$ for two different lockdown strategies. 
    The inset shows the corresponding infected nodes $i(\tau)$ as a function of time. Both plots display $1,000$ curves for both strategies, 
    with a random curve singled out for clarity. }
    \label{fig:CurvesExplainingScanThresh}
\end{figure}

Interesting behaviour emerges as the lockdown threshold is further increased past $\theta_l = 0.05$. 
In contrast to the simple second order phase-transition behaviour when increasing $\lambda$ in the no-lockdown 
case \cite{feld2022:originalsirlargedevpaper}, increasing the lockdown threshold gives rise to rather peculiar behaviour, 
with notable maximum-minimum pairs in the $\bar{C}(\theta_l)$ curve. The positions of these pairs are of interest, 
and are determined by fitting a Gaussian or a log-normal function to $\bar{C}$ near these points. Such pairs are seen 
for $\theta_l = 0.0588$ and $0.0692$, as well as $0.0955$ and $0.1460$. The emergence of these extreme points are most easily 
rationalised by looking at the infection trajectories. In \autoref{fig:CurvesExplainingScanThresh} we show 1,000 such curves $c(\tau)$, as well 
as the corresponding $i(\tau)$ curves, for each $0.0955$ and $0.1460$.

It can be seen that the lockdown is triggered once in both cases. After the lockdown is lifted, the infection curves 
both show a notable second wave of infections. Interestingly $C$ reaches a higher final 
value for $\theta_l = 0.0955$ even though $i(\tau)$ peaks at a higher value for $\theta_l = 0.146$. 
This is due to the fact that an earlier lockdown ensures a larger proportion of the population is susceptible in the second wave, and hence the disease 
can spread to more nodes. This can be seen from the $c(\tau)$ curves, where the earlier lockdown 
leads to a larger cumulative  number of infected nodes in the long-run.

Also note the second peak in the variance around $0.17$. 
This is the transition from the lockdown having some effect at containing the disease to having no effect at all. 

\begin{figure}[htb]
    \centering
    \includegraphics[width =\linewidth]{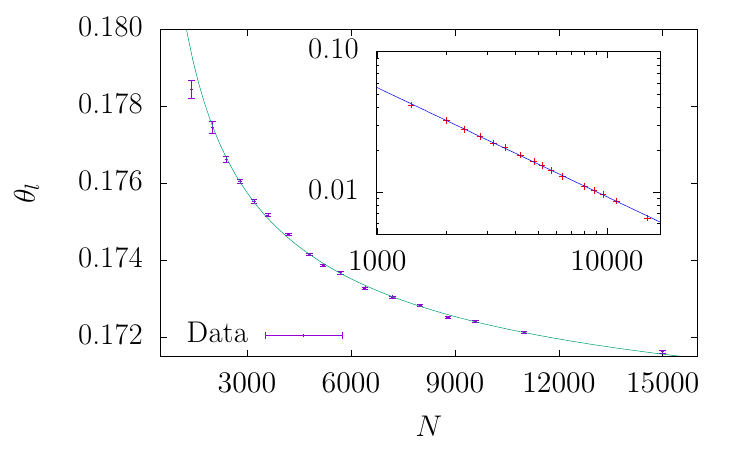}
    \caption{Critical thresholds $\theta_l^c$ for $\eta = 0.586$, $\mu = 0.14$, $\lambda = 0.2$ over the network size $N$. 
    The main plot shows those $\theta_l^c$ corresponding to the transition to the lockdown being ineffective, i.e., 
    the final peak in the variance on \autoref{fig:ScanThresh}. The inset shows those $\theta_l^c$ corresponding to the transition from the 
    lockdown completely containing the disease, i.e., the first peak in the variance in \autoref{fig:ScanThresh}, on a logarithmic scale.}
    \label{fig:CriticalThresh}
\end{figure}

We want to address these parameter-space regions within the
large-deviation simulations later on.  To obtain precise limiting
values of $\theta_l$ for the first and second peak of the variance, we
also performed finite-size scaling. For this purpose, we defined the
finite-size transition points in the  usual way as the peak locations
of the variance $\sigma^2(\theta_l)$. We measured their positions by fitting
Gaussian-shaped functions around the maxima.  
\autoref{fig:CriticalThresh} shows the position of these maxima in $\sigma^2$ as a function of system size. 
The main plot corresponds to the thresholds of the second peak in
$\sigma^2$, while the inset corresponds to the first peak. To actually perform the
finite-size scaling, the
function
\begin{equation}
    \theta_l^c(N) = \theta_l^c(\infty) + a N^{-b}
    \label{eq:power_law_thresh}
\end{equation}
is fitted to the positions of the second maxima, as shown in \autoref{fig:CriticalThresh}, 
giving a value $\theta^c_l(\infty) = 0.1683(5)$ for the critical threshold. 
The other parameters are found to be $a = 0.47(15)$ and $b = 0.52(5)$.

The positions of the variance maxima corresponding to the initial transition are shown on the inset of  \autoref{fig:CriticalThresh} as a 
function of system size. 
The data is nicely linear on a log-log scale, indicating that here the model follows \autoref{eq:power_law_thresh} with $\theta^c_l(\infty) = 0$. 
The other parameters are found to be $a = 12.0(5)$ and $b = 0.778(5)$. 
Thus, this measurement predicts that the threshold capable of completely containing the outbreak is zero for an 
infinitely-sized system. 
We rationalise this by considering the long-range links of the Small-World network. 
With the system nicely in the pandemic phase with $\lambda = 0.2 > 0.1186$, the long range links of the Small-World 
network allow the disease to spread rather quickly from the beginning to all regions of the network. 
For this reason, the spread could only be \emph{prevented} in the thermodynamic limit $N\to\infty$ 
if the system was already in lockdown when the initial five nodes are infected, i.e., $\theta_l = 0$ .

%/Project/data/SimpleSamples/SmallWorld/criticalthresh/gamma0.14/lambda0.2/releaseEnabledFactor8/InsetPlot

\subsection{Disease Duration}\label{sec:disease_duration}
The employed large-deviation algorithm requires a good estimate of the length of the disease outbreak, because
this determines the amount of random numbers that need to be controlled by the Markov chain.
If the duration was chosen to short it would result in too many unfinished outbreak dynamics which in turn would lead
to underestimated values of $C$. 
This would lead to a skewed and incorrectly measured density of states. 
For this reason, before we set up the large-deviation simulations, we investigated 
the life-span of the disease for various system sizes using standard Monte Carlo sampling.

\begin{figure}[htb]
    \centering
    \includegraphics[width=\linewidth]{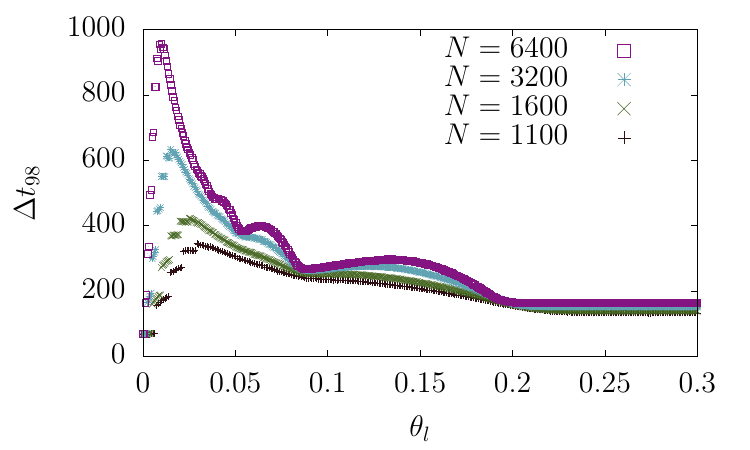}
    \caption{The duration $\tau_{98}$ until $98\%$ of the outbreak simulations reach completion. 
    Note the discontinuities on the left, which are explained in the same manner as those in \autoref{fig:ScanThresh}. }
    \label{fig:lifespanscan}
\end{figure}

For each considered parameter set $(N,\theta_l,\theta_r)$, we measured 
the time $\tau$ it took until $i(\tau) = 0$ was reached.
We did this for $100,000$ networks, respectively, measuring one disease outbreak dynamics each time. 
From this raw data, we extracted the time at which $98\%$ of the outbreaks are completed, 
which we denote by $\tau_{98}(N,\theta_l,\theta_r)$. 
This value is plotted as a function of $\theta_l = 8\theta_r$ for four different 
system sizes in \autoref{fig:lifespanscan}. 

This data is used to decide the duration of the large-deviation simulations as follows. For a given 
parameter set $(N,\theta_l,\theta_r)$, we use a maximum time of  $\tau_{\text{max}} = g \tau_{98}(N,\theta_l,\theta_r)$,
where $g \in [2.7,3.0]$ is a factor that we chose on a case by case basis. 

Our results presented below show that we are able
to sample the distribution $P(C)$ up to a value of $C=1$, i.e., including the
cases where all nodes are infected. This implies that the chosen 
time $\tau_{\text{max}}$ is actually large enough.
Note that we kept track of the duration of the outbreaks encountered during the large-deviation simulations
and counted how often the simulations were \emph{not} finished when we ran out of numbers.
From this we could calculate the frequency $f_{\neq}$ of observing unfinished outbreaks.
For the vast majority of simulations this frequency was $f_{\neq}=0$, i.e., the simulations were long enough.
The worst simulation exhibited a frequency of $f_{\neq}=10^{-5}$, which we still deemed small enough.

\section{Results}
We now present the distributions of the cumulative fraction $C$ of  infected, using the large-deviation methods. 
We firstly show the pdfs $P(C)$. 
We consider the study of the system using large-deviations for the dynamics on a single network. 
This lack of averaging over networks can be justified because
  in a real-life scenario the contact-network is given,
and it is typical. Considering only one, rather large, network per simulation also corresponds 
to assuming  self-averaging. 
Hence we do not consider rare events that occur due to rare networks.

The pdfs are obtained using Wang Landau algorithms and refined with entropic sampling, as explained 
in \autoref{sec:large_Dev_sampling}. Note that all of the results presented below use $\eta = 0.586$, 
$\lambda = 0.2$ and $\mu = 0.14$.

\subsection{Probability Density Functions around the Transition of Lockdown Effectiveness}

\begin{figure}[htb]
    \centering
    \includegraphics[width=\linewidth]{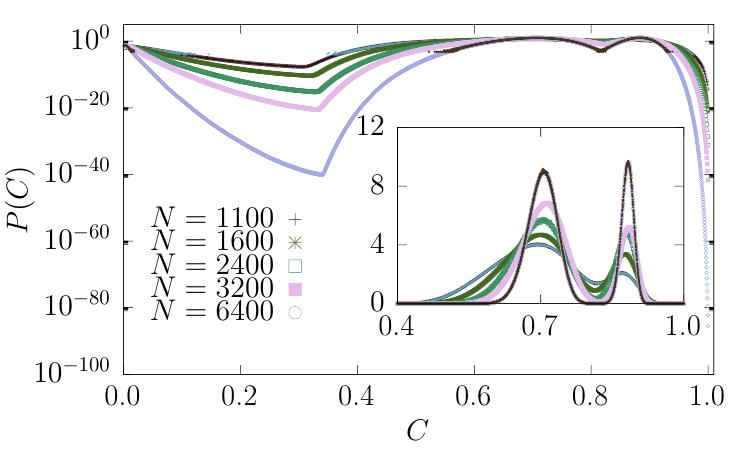}
    \caption{Probability density of the cumulative fraction of infected for several system sizes with $\theta_l = 8\theta_r = 0.1683$, 
    $\eta = 0.586$, $\lambda = 0.2$ and $\mu = 0.14$. 
    The main plot shows the distributions on logarithmic scale, whereas the inset shows the distributions on a linear scale. 
    The system sizes used are $N \in \{1100,1600,2400,3200,6400\}$. The results from standard Monte Carlo sampling are also 
    included for $N = 1100$ and $6400$ with contrasting colour, showing good agreement in the regimes which are
    accessible by such simple sampling.}
    \label{fig:pdfmanysizes}
\end{figure}

Firstly, we observe the pdfs for increasing system size for a lockdown threshold $\theta_l = 8\theta_r = 0.1683$.  
For system sizes below (and including) $N = 1600$, we sample the histograms at the highest possible resolution, i.e., with 
a bin-size of $1/N$. 
For larger system sizes, we increase the bin-size to $2/N$  for computational efficiency. 
The probability densities $P(C)$ for a few system sizes are shown in \autoref{fig:pdfmanysizes}.

We are able to calculate probabilities as low as $10^{-85}$ in the case of $N = 6400$. For this system size, we needed 
to calculated $C$ roughly $10^{9}$ times during the large-deviation simulation. 
Thus, standard Monte Carlo sampling, addressing typical outbreaks,
 would only resolve probabilities of order $\sim 10^{-9}$, 
as shown in the plot where we also included typical sampling with this sample size. 
In the range accessible by typical-event sampling we see a good agreement with the large-deviation data.

The distributions displayed in \autoref{fig:pdfmanysizes} exhibit three peaks. 
One around $C \approx 0$, where the disease quickly dies out, and two for high $C$. 
This pair of peaks appears, as we are right at the transition (variance peak on \autoref{fig:ScanThresh}) 
controlled by lockdown effectiveness $\eta$ in parameter space. 
The system exhibits the highest probabilities for a somehow reduced
 epidemic with $C \approx 0.6$ and
for an unaffected spread with a peak around $C \approx 0.8$.

Notable are also the seemingly non-differentiable points around $C \approx 0.35$ in the pdfs. 
Investigation of the disease trajectories revealed this to be the point where at least one 
lockdown takes place almost for sure. 
This will be discussed in the next section where the number lockdowns as a function of $C$ is also presented.

Further, we connect our model to large-deviation theory by calculating the empirical \emph{rate functions} \cite{den2000large, dembo2009largebook}, 
defined as
\begin{equation}
    \Phi(C) = -\frac{\ln{P(C)}}{N} + \Phi_0,
\end{equation}
where $\Phi_0$ is a shift that ensures each of the $\Phi(C)$ have their minimum at $\Phi = 0$. 

\begin{figure}[htb]
    \centering
    \includegraphics[width=\linewidth]{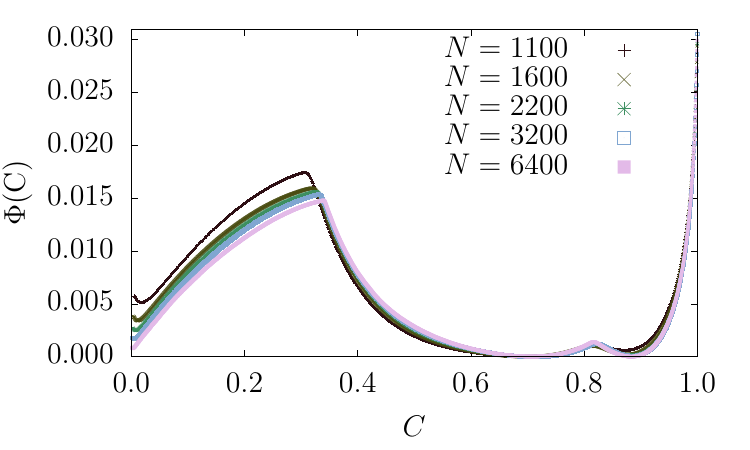}
    \caption{The empirical rate functions $\Phi(C)$ for $N \in \{1100,1600,2400,3200,6400\}$ with $\theta_l = 8\theta_r = 0.1683$ 
    and the other parameters at their default values. }
    \label{fig:RateFunction}
\end{figure}

The obtained rate functions are displayed in \autoref{fig:RateFunction}. 
A convergence with increasing system size is visible, 
 indicating that the large-deviation-principle indeed holds for this model.

 The previous work without lockdowns \cite{feld2022:originalsirlargedevpaper} found this to also be the case, 
 and clearly the lockdowns retain this behaviour.
 Therefore, the pdfs are given to first order by $P(C) \propto \exp\left\{-N\Phi(C) + o(N)\right\}$ with $o(N)$ some sub-linear term. 
 Hence, one could potentially make analytical process in regards to $P(C)$ through application of the
G{\"a}rtner-Ellis theorem \cite{den2000large,touchette2009,dembo2009largebook, touchette2011}, at least
in the region of the convex envelope and where the rate function is differentiable.

\subsection{Parameter Variation}
Secondly, we now investigate the effect of varying the lockdown and release thresholds. 
For different values of $\theta_l$ and $\theta_r$, we calculated and compared the probability density functions. 
The particular values of the parameters correspond to the regions visible in \autoref{fig:ScanThresh},
of which we choose the positions of the minima and maxima of $\bar{C}$. 
Furthermore, we investigated relatively early lockdowns by choosing points where $\bar{C}(\theta_l)$ is
on the first rise. 
For comparison we also considered the critical value of $\theta_l=0.1683$ 
as well as the case with disabled lockdowns. 
The investigated values are presented in \autoref{tab:parameterchoicetable}. 
For actually presenting the pdfs, we have selected a subset of this set for clarity.

\begin{table}[htb]
  \centering
    \begin{tabular}{|c|c|}
    \hline
    $\theta_l$  & $\theta_r$ \\
    \hline
    0.0105 & 0.0013 \\
    0.0210 & 0.0026 \\
    0.0421 & 0.0053 \\
    0.0588 & 0.0074 \\
    0.0692 & 0.0087 \\
    0.0955 & 0.0119 \\
    0.1460 & 0.0183 \\
    0.1683 & 0.0210 \\
    Disabled & Disabled \\
    \hline
    \end{tabular}%
      \caption{Points of interest in parameter space.}
  \label{tab:parameterchoicetable}%
\end{table}%

With these values in mind, we ideally should only vary one parameter at a time. 
For this reason, we first fixed the lockdown threshold at the critical value of $0.1683$, and varied
 the release threshold $\theta_r$ according to the values presented in \autoref{tab:parameterchoicetable}, 
where we also include a simulation where releasing is disabled completely, i.e., non-lifted
 lockdowns. 
This is the subject of \autoref{sec:subsubrelease}. 
Secondly, we fix the release-threshold at $0.0210$ and vary the lockdown threshold according to the (admissible) values 
from \autoref{tab:parameterchoicetable}. 
Finally, we `compile' these results together by investigating the pdfs of the constant ratio $\theta_l = 8\theta_r$ in \autoref{sec:subsubboth}.

\subsubsection{Varying the Release Threshold}\label{sec:subsubrelease}

\begin{figure}[htb]
    \centering
    \includegraphics[width=\linewidth]{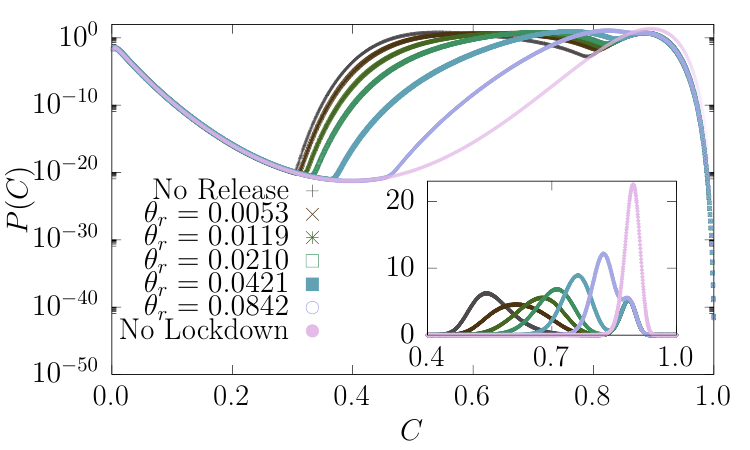}
    \caption{The probability density functions $P(C)$ for varying the release threshold, with the lockdown threshold fixed at $\theta_l = 0.1683$. 
    The remaining parameters are $N = 3200$, $\mu = 0.14$, $\lambda = 0.2$ and $\eta = 0.586$. 
    The main plot is the entire pdf on a logarithmic scale, while the linear scale is shown on the inset.   
    The values of $P(C < 0.4)$ are omitted on the linear scale as they are practically invisible here.
    }
    \label{fig:VaryReleasePlot}
\end{figure}

Here we study the effect on the pdfs of varying the release threshold. The pdfs are shown on  \autoref{fig:VaryReleasePlot}. 
It can be seen that the distributions actually align with the no-lockdown case before deviating away at a 
particular $C$ value specific to each parameter set. 
These are the $C$ values where the lockdowns become relevant for that particular parameter. 
The first to deviate is the $\theta_r = \text{disabled} \equiv 0$ case, 
with the point of deviation, also marked by a non-analyticity, visibly increasing with $\theta_r$. 

This can be explained as follows. If the lockdown is released the disease will likely be able to propagate 
through the system better and therefore infect more people compared to the case where the lockdown is not lifted.
For rather larger values of $C$ this results in a higher probability, and, due to normalization,
for medium values of $C$ in a lower probability. For very small values of $C$, the lockdown is never triggered. 

Interestingly, for all cases the non-analytic point where the deviation between the pdfs for lockdown and
not lockdown appears is considerably larger, at least $C\ge 0.3$,
 than the lockdown threshold $\theta_l=0.1683$ here!
This means, the location of this points is not only determined by the value of $\theta_l$ but also
by the complex network topology.

Furthermore, it can be seen that increasing the release threshold from zero deforms the pdf by increasing the height of the first peak at 
intermediate $C$, with alignment around the second peak, i.e. for $C > 0.85$, indicating the behaviour at 
high $C$ is independent of the release threshold. 
We then deduce the behaviour of the pdf is, for large $C$, likely dictated by the lockdown threshold. 

\begin{figure}
    \centering
    \includegraphics[width=\linewidth]{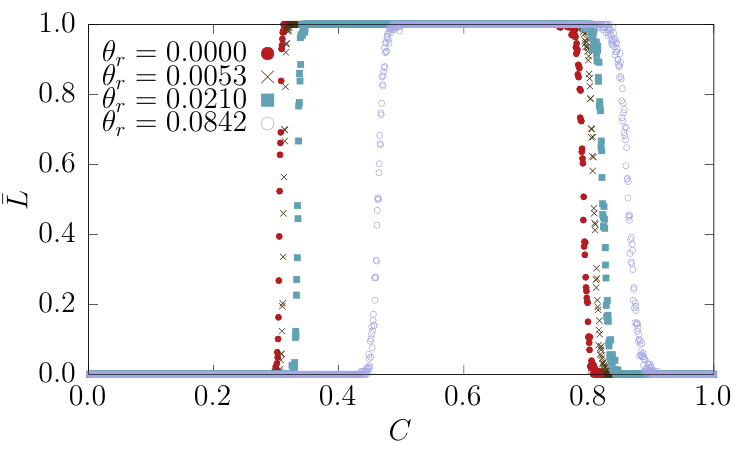}
    \caption{The average number $\bar{L}$ 
    of lockdowns as a function of $C$ for varying the release threshold with $\theta_l = 0.1683$ fixed. 
    The remaining parameters are $N = 3200, \mu = 0.14, \lambda = 0.2$ and $\eta = 0.586$.}
    \label{fig:varyrelease:lockdownplot}
\end{figure}

This is emphasised by investigating the average number $\bar{L}$ 
of lockdowns, see Figure \ref{fig:varyrelease:lockdownplot}.
Interestingly, for high values of $C$, lockdowns were not triggered. This means they exhibit values for the numbers
of infected below the lockdown threshold, i.e., 
extremely severe outbreaks are slow-spreading. 
Also, for the present case with the lockdown threshold fixed at $\theta_l = 0.1683$,
 the system was found to exhibit typically at most a single lockdown. 
Note the appearance of a transition between zero and one lockdown,
 This transition corresponds very well
to the nonanalytic point seen in \autoref{fig:VaryReleasePlot} and the
 transition point increases monotonically with $\theta_r$.

This can be explained as follows. To reach the 
lockdown threshold,  the system needs a substantial amount of simultaneously infected nodes.
If the lockdown is then quickly lifted due to high value of $\theta_r$
the resulting value of $C$ will be larger then when 
the lockdown is lifted late $\theta_r \to 0$. Thus, the $\bar{L}(C)$ curves are shifted to the right.
Note that $C$ is not an independent parameter, i.e., cannot be directly controlled,
 although it is plotted on the $x$-axis here!
For lower values of $\theta_r$ the system spends more time in lockdown where the disease spread is greatly limited.
 Thus,  the only way for the system to achieve higher value of $C$ is to have never locked down in the first place, 
 which can only occur inspite of large values of $C$ if the spreading is slow.

\subsubsection{Varying the Lockdown Threshold}\label{sec:subsublock}

\begin{figure}[htb]
    \centering
    \includegraphics[width=\linewidth]{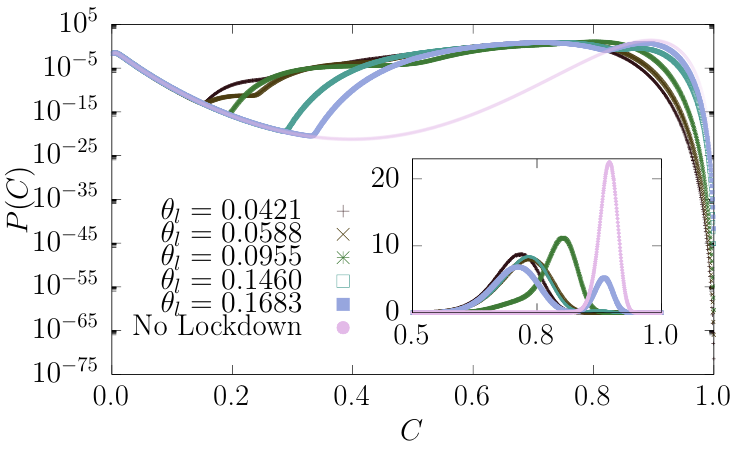}
    \caption{The probability density function $P(C)$ for varying the lockdown threshold, with $\theta_r = 0.0210$ fixed. The remaining parameters are
    $N = 3200$, $\mu = 0.14$, $\lambda = 0.2$ and $\eta = 0.586$. As before, only the visible portion of the linear scale is shown.   }
    \label{fig:VaryLockPlot}
\end{figure}

The pdfs $P(C)$ for varying the lockdown threshold $\theta_l$, are shown in 
\autoref{fig:VaryLockPlot}.  
Note that the release threshold is fixed at $\theta_r = 0.0210$. 
As before, normalisation ensures the pdfs align with the no-lockdown 
case for small values of $C$, because as long as $i(\tau) \leq C$ holds no
lockdown can be triggered.
The curves start deviating at a value of $C$ that here depends on $\theta_l$,
again accompanied by a non-analyticity of $P(C)$. 
These points exhibit values of $C$ that are clearly larger than the respective values of $\theta_l$. 
It is worth mentioning that for low values of $\theta_l$ more than one single non-analytic point appears.

Also, increasing the lockdown threshold mainly moves the dominant peak to the right. 
However, with increasing lockdown threshold, we approach the case where the lockdowns fail to affect the 
system as the threshold is never reached. 
This gives rise to a small second peak at first, which is visible in the pdf for $\theta_l=0.146$ in the logarithmic scale. 
At the critical threshold $\theta_l=0.1683$ the second peak is of notable magnitude and also visible in the linear scale.
Finally, with disabled lockdown, the once dominant peak has vanished completely and only this second 
peak remains, apart from the peak near $C=0$ of course.

\begin{figure}[htb]
    \centering
    \includegraphics[width=\linewidth]{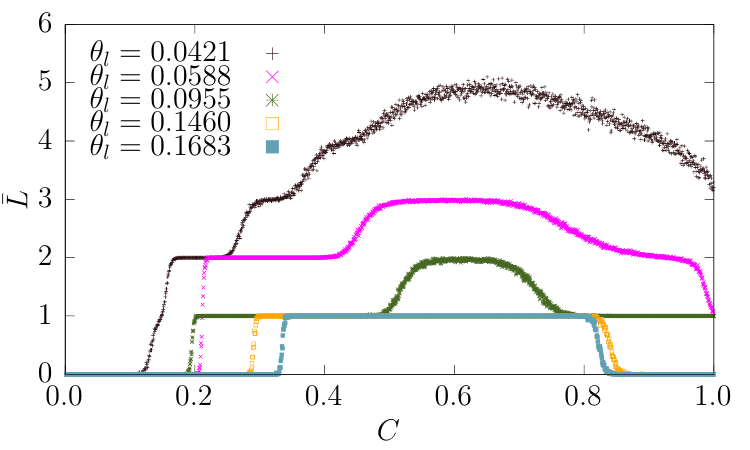}
    \caption{The average number $\bar{L}$ of lockdowns as a function of $C$ for varying the lockdown threshold with $\theta_r = 0.0210$ fixed. 
    The remaining parameters are
    $N = 3200$, $\mu = 0.14$, $\lambda = 0.2$ and $\eta = 0.586$. }
    \label{fig:varylock:lockdownsplot}
\end{figure}

We also notice that in contrast to the  previous case of varying the release threshold, varying the
lockdown threshold \emph{is} still having a strong impact on the probabilities even for $C > 0.85$. 
In particular the behaviour gets richer because  low lockdown thresholds result in triggering several
 lockdown-release pairs. This can be seen in  \autoref{fig:varylock:lockdownsplot},
where we display the average number $\bar{L}$  of triggered lockdowns
 as function of $C$.
Again, the points where $\bar{L}$ changes strongly correspond to the non-analytic points of $P(C)$.
Indeed a system with a low lockdown threshold must undergo multiple lockdowns in order to attain high $C$ values. 
By contrast, higher lockdown thresholds allow the disease to spread a bit more rapidly,
although to obtain large $C$ values it must still spread slow enough to stay below the lockdown threshold. 
It is also seen that intermediate thresholds can still suffer a severe outbreak with a single lockdown on average.

\subsubsection{Fixed-ratio variation}\label{sec:subsubboth}

\begin{figure}[htb]
    \centering
    \includegraphics[width=\linewidth]{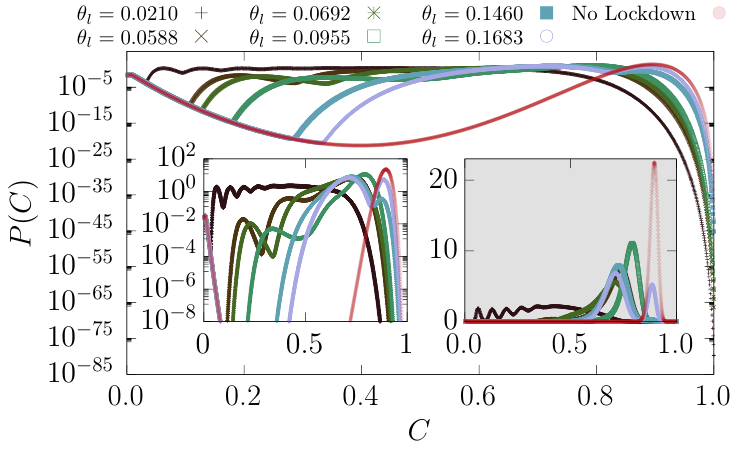}
    \caption{Probability density of the the cumulative fraction of infected for $N = 3200$, $\mu = 0.14$, $\lambda = 0.2$ and $\eta = 0.586$. 
    The lockdown and release threshold are varied simultaneously but constrained to the ratio $\theta_l = 8\theta_r$. 
    The left inset (white background) restricts the probability range to $P(C) \geq 10^{-8}$ to show the finer detail in this regime. 
    The right inset (darker background) shows the distributions with 
    linear scale.}
    \label{fig:VaryBothPlot}
\end{figure}

In \autoref{fig:VaryBothPlot} we investigate how the probability distribution functions change with the lockdown and release 
thresholds simultaneously varied, 
but with their ratio kept at a constant factor of eight.

\begin{comment}
\begin{figure}[htb]
    \centering
    \includegraphics[width=\linewidth]{Figures/varyboth_inset.pdf}
    \caption{Probability density of the the cumulative fraction of infected for $N = 3200$, $\mu = 0.14$, $\lambda = 0.2$ and $\eta = 0.586$. 
    The lockdown and release threshold are varied simultaneously but constrained to the ratio $\theta_l = 8\theta_r$.}
    \label{fig:VaryBothPlot}
\end{figure}
\end{comment}

The left part of the curves aligns with the no-lockdown curve,
 for the  same reasons as discussed before.
For low values of $\theta_l$, i.e.,  quickly triggered lockdowns,
the system heavily favours 
low value of $C$, with probabilities as small as $10^{-80}$ for high values $C$. 
Increasing the lockdown threshold of course allows the system to exhibit 
an increasingly higher fraction $C$ of infected.

With respect to the pdf's shape, the result for the lowest threshold, 
i.e., $\theta_l=0.021$ ,  exhibits multiple peaks of similar heights. 
This is due to multiple lockdowns and is discussed below. 
Note that this value of $\theta_l$ is close to the peak location 
of the variance, which corresponds to the epidemic threshold, 
see  \autoref{fig:ScanThresh}.

Increasing the threshold beyond the critical threshold shifts the behaviour of the system such that it is in a strong epidemic phase.
Most of the peaks of the pdf at small values of $C$ disappear 
somewhere between $\theta_l= 0.021$ and $\theta_l=0.0588$.

%For $\theta_l=0.0588$ we see a peak around $C=0.4$ that disappears 
%when increasing the threshold,
%as the slope of the distribution changes,
%whereas the peak around $C=0.2$ first shifts to the right before disappearing completely.

Still, most of the pdfs also exhibit a peak around $C=0.7$,
which corresponds to a rather large outbreak, that is  
affected by the lockdowns.
A notable exception for this is $\theta_l=0.0955$, where this peak is slightly 
shifted  a bit to higher values of $C$, i.e., here this typical outbreak is 
larger.
Starting with $\theta_l=0.146$ we see a peak around $C=0.9$ emerging, 
which corresponds 
to the peak without lockdowns. Thus, the lockdown threshold for these systems
 is so high that the high values of $C$ are reached without triggering the lockdown anymore.

\begin{figure}[htb]
    \centering
    \includegraphics[width=\linewidth]{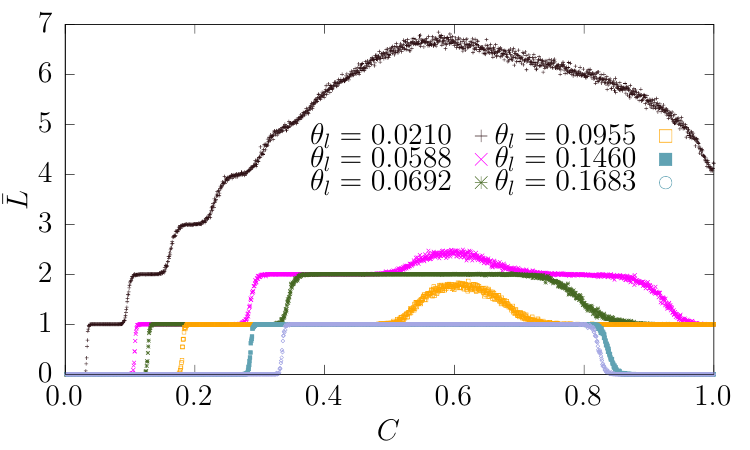}
    \caption{The average number of lockdowns as a function of $C$ for varying the lockdown and release thresholds simultaneously 
    with $\theta_l = 8\theta_r$. The remaining parameters are $N = 3200$, $\mu = 0.14$, $\lambda = 0.2$ and $\eta = 0.586$.}
    \label{fig:numlockdownsbothvarying}
\end{figure}

To understand the shapes of the pdfs better, we look at the 
average number $\bar{L}$ of lockdowns as shown in
 \autoref{fig:numlockdownsbothvarying}. Indeed, not more
than one lockdown is triggered for $\theta_l\geq 0.146$.

Consistently with the two previous cases, low $C$ values exhibit 
relatively few lockdowns because the disease goes extinct rather early. 
Still, the case of the relatively low lockdown threshold $\theta_l=0.021$ 
creates multiple lockdowns, i.e., several infection waves.

As was the case previously, the maximum number of lockdowns typically occurs around intermediate values of $C$.
For high values of $C$ the slowly spreading infections dominate the dynamics,
which leads  to relatively few lockdowns again. For 
high threshold values, even no lockdowns are triggered there.

To understand what is going on for the lowest considered threshold 
value $\theta_l$ in more detail we 
now look at the conditional probability $P(L|C)$ that 
$L$ lockdowns are triggered given 
a value of $C$. The normalization is such that $\sum_L P(L|C)=1$ for all
values of $C$.

\begin{figure}[htb]
    \centering
    \includegraphics[width=\linewidth]{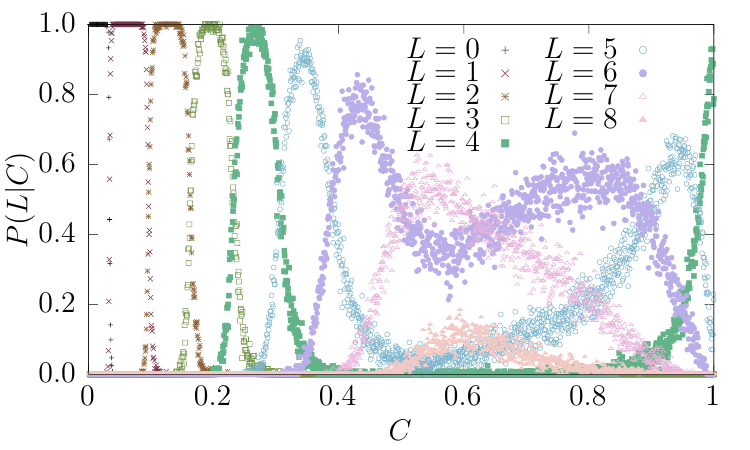}
    \caption{The conditional probability $P(L|C)$ that the system exhibits
    $L$ lockdowns given a specific value of $C$.
    The parameters are $N = 3200$, $\mu = 0.14$, $\lambda = 0.2$, $\eta = 0.586$, $\theta_l=0.021$ and $\theta_r = \theta_l/8$. 
    }
    \label{fig:lockdownsboth}
\end{figure}

Interestingly we can see that for about  $C\le 0.45$  the dynamic
is dominated by a specific number $L$. 
In contrast, for values around $C \approx 0.6$,  the distributions
exhibit has a notable probability 
for 6, 7 or 8 lockdowns. This region is also characterized by
large fluctuations.
For even larger values of $C$, the typical number of lockdowns decreases again.

\subsection{Correlation and Heatmaps}

\begin{figure}[htb]
    \centering
    \includegraphics[width = \linewidth]{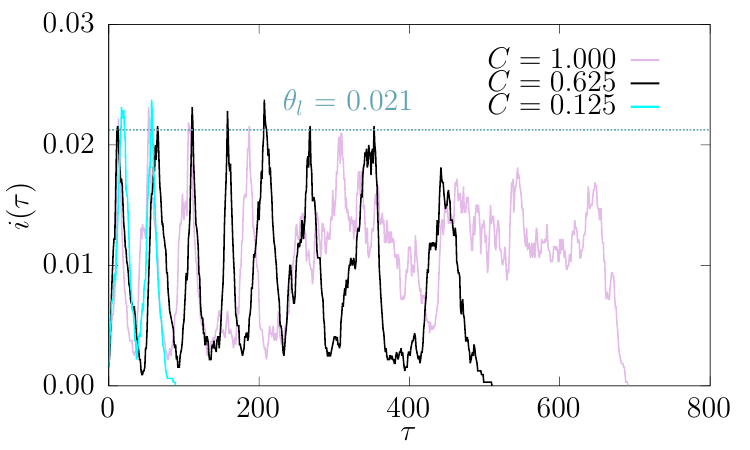}
    \caption{Examples of infection time series $i(\tau)$ 
for $N=3200$, $\theta_l = 0.0201$ , 
$\theta_r=\theta_l/8$ and 
different values of $C$, respectively. 
    The dashed horizontal line indicates the lockdown threshold. }
    \label{fig:curves}
\end{figure}

To analyse the outbreak dynamics further \cite{feld2022:originalsirlargedevpaper}, we store a number of the outbreak trajectories 
during the entropic sampling. We elected to store $200,000$ such curves for each pdf. The trajectories are binned according to their corresponding
value of  $C$. Let $T$ denote a time series, $T=i$ or $T=c$, with 
\begin{equation}
    T = (T(0), \dots T(\tau_{\text{max}}-1)).
\end{equation}
Three examples of such curves are shown on Figure \ref{fig:curves} for $\theta_l = 8\theta_r = 0.021$ and for low, medium and high value of $C$, respectively. 
They present the general behaviour in each of the three regimes. 
For low values of $C$, typically very short-lived trajectories with 
one or two lockdowns appear, as seen as the previous section.
For intermediate values of $C$ one may experience many lockdowns.
For high $C$ typically only a few or no lockdowns occur, depending of 
course on the parameter set. 

In the following  subsection we use these time series to construct heatmaps 
to investigate the similarity of pairs of time series. 
Secondly, heatmaps pertaining to other properties of the outbreaks are 
presented and discussed.

\subsubsection{Disparity Maps}

To compare the similarity of two outbreaks, i.e., of two time series $T_1$ and $T_2$, we first normalise by their respective maxima. 
The length of the time series is denoted by $l_1$ and $l_2$ respectively.
Let now $l_{\max} = \max\{l_1,l_2\}$. A distance $d$ between two time series is defined \cite{feld2022:originalsirlargedevpaper} as
\begin{equation}
    d(T_1,T_2) = \frac{1}{l_{\text{max}}} \sum_{\tau = 0}^{l_{\max}-1} \left|T_1(\tau)-T_2(\tau)\right|\,.
\end{equation}
The disparity $V_T(C_1,C_2)$ is the averaged distance for all pairs of 
time series $T$ with bin 
values $C_1$, $C_2$, respectively. 

\begin{figure}[htb]
    \centering
    \includegraphics[width = \linewidth]{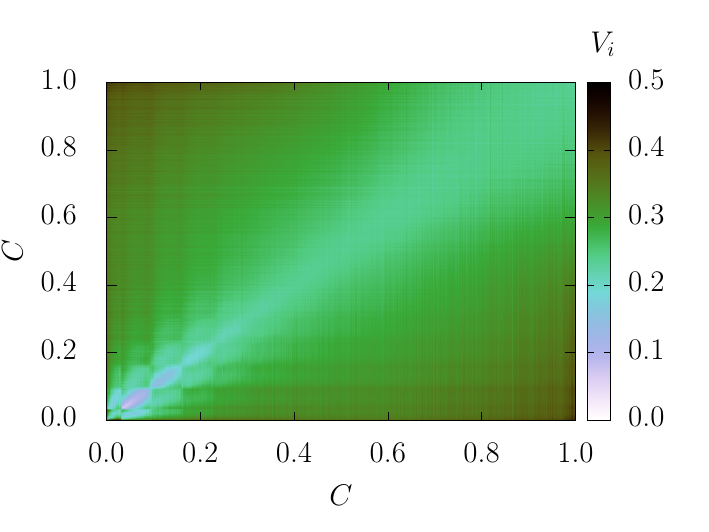}
    \caption{Disparity $V_i$ of the infection curves $i(\tau)$ for $\theta_l = 8\theta_r = 0.0210$. 
    The curves have been binned with respect to their $C$ value, with $N = 3200, \lambda = 0.2, \mu = 0.14$ and $\eta = 0.586$. 
    }
    \label{fig:l0.0210r0.0026infecteddisparity}
\end{figure}

For brevity, we only present the disparity $V_i$ of the infection curves 
$i(\tau)$ here.
Figure \ref{fig:l0.0210r0.0026infecteddisparity} shows $V_i$ 
as a heatmap for $\theta_l = 8\theta_r = 0.0210$. 
Note that using the large-deviation approach has allowed us to create such a 
heatmap over the $\emph{entire}$ allowed range of $C$; 
particularly also in the range which is inaccessible by 
standard Monte Carlo sampling. 

The disparity appears to form regions  in the heatmap. 
Looking at the diagonal, which represents comparing the curves in one bin to one another, for $0<C<0.2$ the curves are evidently rather similar. 
These low-outbreak curves are characterized by early lockdowns stopping the spread of the disease. 
Following the diagonal further the disparity increases and one can no longer 
clearly distinguish regions from one another.
This also makes sense as an increasing number of lockdowns makes it 
increasingly 
unlikely that the lockdowns  of two infection curves occur at the same 
time step. Since 
the lockdowns trigger rapid changes in the time evolution of the disease
 this increases
the disparity between two such time series.

Comparing the time series of any fixed $C$ value, e.g., $C=0.2$ with the 
other time series we can see that they quickly become dissimilar 
to one another when the two corresponding values of $C$ differ.
Looking at \autoref{fig:lockdownsboth} this makes sense,
as the number of lockdowns triggered varies a lot, but has a high 
correlation with the $C$ value. 
Also the visual ``steps'' in similarity in the range of small values of $C$
correspond to changes in the typical number $L$ of lockdowns.

\begin{figure}[htb]
    \centering
    \includegraphics[width = \linewidth]{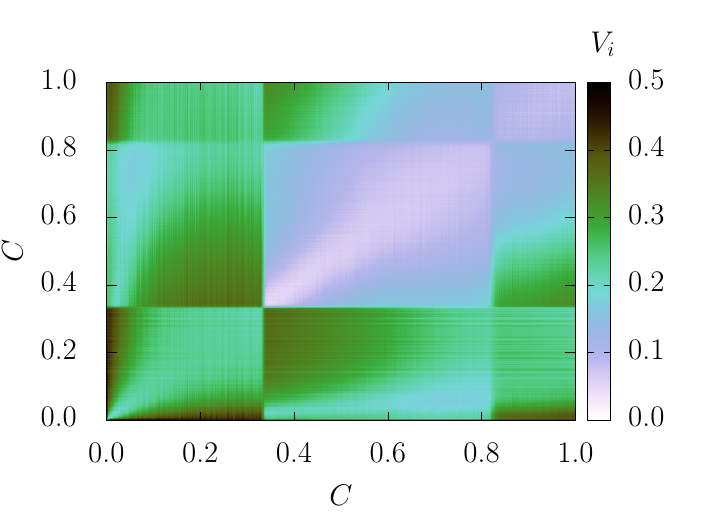}
    \caption{Disparity $V_i$ of the infection curves $i(\tau)$ for $\theta_l = 8\theta_r = 0.1683$. 
    The curves have been binned with respect to their $C$ value, with $N = 3200, \lambda = 0.2, \mu = 0.14$ and $\eta = 0.586$.
    }
    \label{fig:l0.1683r0.0210infecteddisparity}
\end{figure}

We also show the disparity $V_i$ for a higher value 
$\theta_l = 8\theta_r = 0.1683$, around the critical threshold, in \autoref{fig:l0.1683r0.0210infecteddisparity}. 
Here, three regions can be distinguished. 
Looking at \autoref{fig:numlockdownsbothvarying} we see that the first region
$0\leq C \leq 0.34$ corresponds to no lockdowns,
whereas $0.34 < C \leq 0.81$ corresponds to 1 lockdown and the last region
$C\geq 0.81$ exhibits no lockdowns again.
Note that the border of region one and two corresponds to the 
seemingly non-analytic behaviour seen on the pdf in \autoref{fig:VaryBothPlot}.

The time series in the region of $0.34<C<0.81$ exhibit extremely
 low disparities to one another, which shows that there is low variability 
in the time series. 
This is also true, although to a slightly less degree, 
for the third region, i.e., $C\geq 0.81$. Interestingly, also off-diagonal
parts representing the disparity between time series from different 
regions exhibit an internal structure to some degree. This indicates
that the three regions are somehow subdivided further. We do not go into details
here.

\subsubsection{Conditional Density}

Other properties of the outbreaks can be studied by considering the conditional densities $\rho(Q|C)=P(Q,C)/P(C)$, with $Q$ some measurable quantity. 
The time series are binned according to the cumulative fraction $C$ of infections, and then a histogram of $Q$ given $C$ is constructed. 
These are presented as heatmaps. 
In this paper, we consider the following quantities for $Q$: 
\begin{itemize}
    \item $Q = \tau_{\max}$, that is the number of time steps for the infection trajectories to reach their 
    \emph{global} maximum of $i(\tau)$. If a trajectory reaches the same maximum multiple times we take the time it took to reach the first one,
    \item $Q = M$, the relative hight of the global maximum, i.e., $i(\tau_{\max})$.
\end{itemize}

Across the various data sets, some similar patterns emerge in the heatmaps. 
To illustrate the main points, we only present those for $\theta_l = 8\theta_r = 0.0421$ and $0.1683$ respectively.

\begin{figure}
    \centering
    \includegraphics[width = 0.45\textwidth]{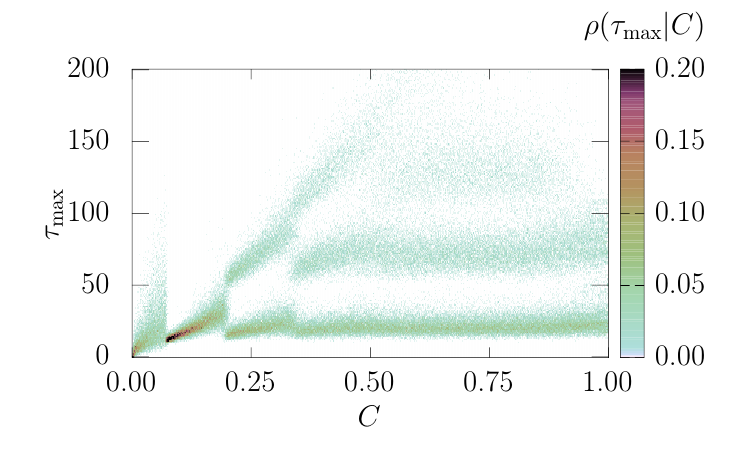}
    \caption{The conditional density $\rho(C|\tau_{\text{max}})$ for $\theta_l = 8\theta_r = 0.0421$. 
    This is the amount of time steps the system requires to reach its global maximum in the infection curves $i(\tau)$ for each $C$ value. 
    The other parameters are $N = 3200, \lambda = 0.2, \mu = 0.14$ and $\eta = 0.586$.}

    \label{fig:indexmax}
\end{figure}

The conditional density $\rho(\tau_{\text{max}}|C)$ for $\theta_l=0.0421$ is shown on \autoref{fig:indexmax}. 
As we saw previously, with these values for the parameters,
 the system has a tendency to experience multiple lockdowns. 
The lockdowns will lead to a steep decline in $i(\tau)$. Therefore, 
if lockdowns occur, the \emph{global} maximum will be very close to
one of the times where the lockdown was triggered, which can be at different times,
but typically not at all times.
This is revealed in this heatmap by the multiple `bands' of high probability
which are visible for large values of $C$. 
Note that with the propagation of the disease, with increasing time,
less and less susceptible nodes remain. Therefore,
 the earlier lockdowns have a higher chance to
lead to the global maximum, which is apparent by the color of 
the earlier `bands'.

There are multiple discontinuities in $P(\tau_{\max}|C)$. They, like above, 
correspond to the $C$ values where the 
dominant number of lockdowns changes.

\begin{figure}
    \centering
    \includegraphics[width=\linewidth]{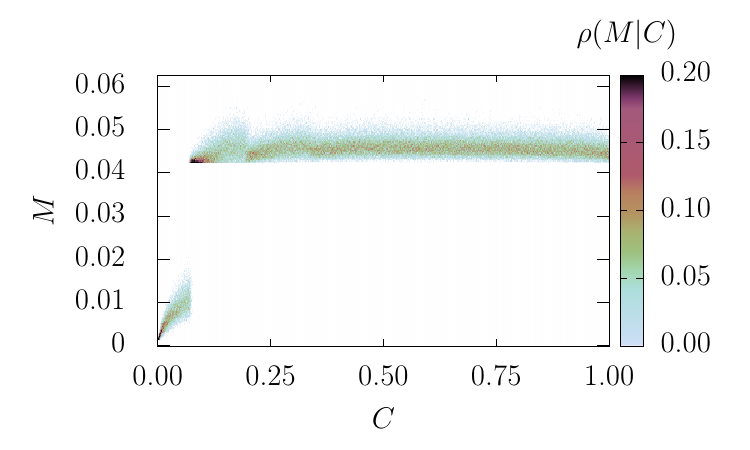}
    \caption{The conditional density $\rho(M|C)$ for $\theta_l = 8\theta_r = 0.0421$. 
    Here $M$ is the maximum fraction of the network that was simultaneously infected. 
    The other parameters are $N = 3200, \lambda = 0.2, \mu = 0.14$ and $\eta = 0.586$.}
    \label{fig:valmax}
\end{figure}

\autoref{fig:valmax} shows the conditional density $\rho(M|C)=\rho(M,C)/P(C)$ 
for $\theta_l = 8\theta_r = 0.0421$. 
Note the phase-transition like discontinuity for $C \approx 0.05$, 
 where the maximum $M$ is somehow constrained by the lockdown threshold. 
For low values of $C$ clearly no lockdowns are triggered. When $C$ exceeds
 $\theta_l$ significantly, lockdowns are triggered and the number of
infections is drastically reduced. This leads to a significant increase
of $P(C)$ for values of $C$ just above $\theta_l$, as we saw before
and to a significant increase of $P(M,C\ge \theta_l)$ 
for values of $M$ also just above
of $\theta_l$, i.e., a peak near $M=0.045$.  
Note that in $P(C,M)$ also the small peak
for $M$ near 0.01 continues to exist for $C$ larger than $\theta_l$, 
but due to the normalization
by a much larger value of $P(C)$, as compared to $C<\Theta_l$, 
the small peak is not visible any more in \autoref{fig:valmax}. Thus,
the other peak  is dominant for $C$ larger than $\theta_l$
and appears as a discontinuity.

By contrast, to attain higher $C$ the infection curves must spread with more vigour, triggering the lockdowns and having their 
global maximum restricted by the lockdown threshold.

\begin{figure}
    \centering
    \includegraphics[width=\linewidth]{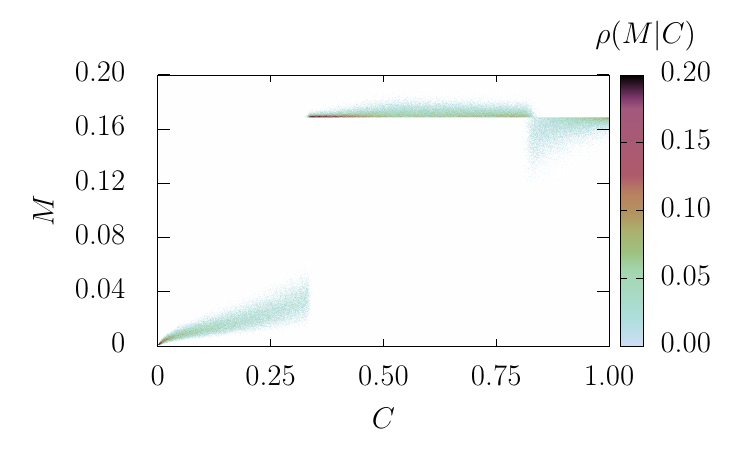}
    \caption{The conditional density $\rho(M|C)$ for $\theta_l = 8\theta_r = 0.1683$. 
    Here $M$ is the maximum fraction of the network that was simultaneously infected. 
    The other parameters are $N = 3200, \lambda = 0.2, \mu = 0.14$ and $\eta = 0.586$.}
    \label{fig:valmax0.1683}
\end{figure}

On this note, with higher lockdown threshold a third phase in the $\rho(M|C)$ surface begins to appear. 
This is shown on \autoref{fig:valmax0.1683} for $\theta_l = 8\theta_r = 0.1683$.

It can be seen that the behaviour is qualitatively similar to that of \autoref{fig:valmax}, with the emergence of a new phase around $C > 0.8$. 
This is because the relatively high lockdown threshold allows for slowly 
spreading but long-lasting trajectories with relatively high value of $M$
 yet still 
lower than the lockdown threshold. 
This is seen from the non-zero probability for $M$ lower than the 
lockdown threshold in this regime.

\section{Summary and Concluding Remarks}

We have studied a stochastic SIR network model under the influence of 
lockdowns. 
We employed an infection-level activated lockdown where the lockdown was
implemented by temporarily removing a certain percentage of the edges in 
the Small-World network. 
The goal was to obtain the complete density of states of the fraction $C$ 
of infected nodes for a variety of lockdown and release thresholds. 

The parameter sets of interest were chosen by using regular 
infection dynamics  simulations, where critical transition thresholds 
where found separating phases where the lockdown was
 effective or not, respectively. 
The values of interest for the lockdown thresholds $\theta_l$ 
were chosen by considering interesting points, maxima and minima,
 of the curves for the average $C$ as a function of $\theta_l$. 
The severity of the lockdowns, that is the fraction of edges to be removed, 
was taken as the percolation threshold of the particular Small-World 
networks employed.

The density of states were obtained using a Wang-Landau algorithm with refinement via entropic sampling. 
Probability densities as small as $10^{-85}$ were obtained in this fashion. 
Furthermore, rate functions were calculated which showed consistency with the 
large-deviation principle, which means that $P(C)$ falls into a standard
class of behaviour. 
In particular we observed the appearance of nonanalytic points 
of $P(C)$ which are not present in the no-lockdown case. 

The shapes of the pdfs were rationalised by analysing
 the infection trajectories. It was found outbreaks exhibiting a low value of
$C$ either  die out almost instantly, or trigger the lockdown, 
sometimes multiple times, before becoming extinct. 
For intermediate values of $C$ outbreaks were seen to spread violently 
leading typically to several lockdowns. 
Finally, outbreaks with high value of $C$  typically exhibit  
slowly-developing dynamics
 with few to none triggering of a lockdown. 

The disparity heatmaps further reflected this kind of behaviour, with some showing discontinuous changes between regimes. 
Moreover the tendency of the system to exhibit several lockdowns 
with discontinuous transitions was seen in the behaviour 
of conditional probability densities obtained from the trajectories.

For practical applications, it should be stressed that we observed
at least two types of pandemic outbreaks. First, the short but heavy ones, 
which triggered one or several lockdowns. On the other hand, there are
strong but slowly-developing outbreaks, where a lockdown has never
been triggered.
While the probability of the latter was rather low, especially for small lockdown 
thresholds, this effect could increase when other factors are also included, e.g., a latent period 
of the disease as done in the SEIR model \cite{yan2019}.

For public health control this means that, if the goal lies not only in minimizing $M$ but also in minimizing $C$, that one should
not only look at the current number of infected individuals but try to find
other criteria to issue lockdowns or consider different measures altogether. These criteria will likely depend
on more complex analyses of the state of a network and could involve
the size of the infection front, i.e., the actual active 
contacts between infected and susceptible individuals.

In the future we plan to study in a similar fashion
 the transfer of diseases between animals and humans, i.e., zoonoses. 
Such a transfer is in general
not highly probable, 
at least for those infections where the transfer to humans
has not taken place yet. Thus, the application of the large-deviation
approach will be very useful here, building upon the expertise
we have gathered so far for the  one-species model.

\section*{Acknowledgements}

We thank the German Academic Exchange Service (DAAD) for supporting L.~P.~Mulholland trough 
the RISE program and thereby partially funding this collaboration. 

Y.~Feld has been financially supported by the German Academic Scholarship Foundation (Studienstiftung des Deutschen Volkes). 

The simulations were performed at the HPC Cluster CARL, located at the University of Oldenburg
(Germany) and funded by the DFG through its Major Research Instrumentation Program (INST 184/157-1 FUGG) and the Ministry 
of Science and Culture (MWK) of the Lower Saxony State.

\bibliography{bibliography}
\end{document}